\documentclass[conference]{IEEEtran}
\IEEEoverridecommandlockouts

\usepackage{cite}
\usepackage{amsmath,amssymb,amsfonts}
\usepackage{graphicx}
\usepackage{textcomp}
\usepackage{xcolor}
\usepackage{subcaption}
\usepackage[linesnumbered,ruled,vlined]{algorithm2e}
\usepackage{blindtext}
\usepackage{tabulary}

\begin{document}

\title{Boosting Communication Efficiency of Federated Learning's Secure Aggregation}

\author{
\IEEEauthorblockN{Niousha~Nazemi\textsuperscript{1},
    Omid~Tavallaie\textsuperscript{1},
    Shuaijun~Chen\textsuperscript{1},   
    Albert~Y.~Zomaya\textsuperscript{1},
    Ralph~Holz\textsuperscript{1,2}
   }
\IEEEauthorblockA{
    \hspace{1cm}\textsuperscript{1}School of Computer Science, The University of Sydney, Australia\\
    \textsuperscript{2}Faculty of Mathematics and Computer Science, University of Münster, Germany \\
    \{niousha.nazemi, omid.tavallaie, albert.zomaya, ralph.holz\}@sydney.edu.au, sche5840@uni.sydney.edu.au}
\\
}

\maketitle

\begin{abstract}
Federated Learning (FL) is a decentralized machine learning approach where client devices train models locally and send them to a server that performs aggregation to generate a global model. FL is vulnerable to model inversion attacks, where the server can infer sensitive client data from trained models. Google's Secure Aggregation (SecAgg) protocol addresses this data privacy issue by masking each client's trained model using shared secrets and individual elements generated locally on the client's device. Although SecAgg effectively preserves privacy, it imposes considerable communication and computation overhead, especially as network size increases. Building upon SecAgg, this poster introduces a Communication-Efficient Secure Aggregation (CESA) protocol that substantially reduces this overhead by using only two shared secrets per client to mask the model. We propose our method for stable networks with low delay variation and limited client dropouts. CESA is independent of the data distribution and network size (for higher than 6 nodes), preventing the honest-but-curious server from accessing unmasked models. Our initial evaluation reveals that CESA significantly reduces the communication cost compared to SecAgg.
\end{abstract}


\section{Introduction}
\vspace{-0.15mm}

\begin{figure}[t]
  \centering
    \includegraphics[width=1\linewidth , height=30 mm]{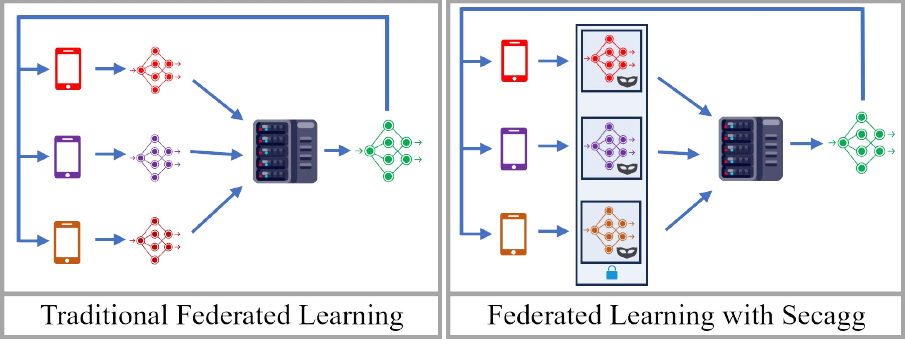}
  \caption{Comparison between Vanilla FL and FL with SecAgg.}
  \label{fig:comparison_fl_secagg}
\end{figure}

\begin{figure}[t]
  \centering
    \includegraphics[width=75 mm, height=30 mm]{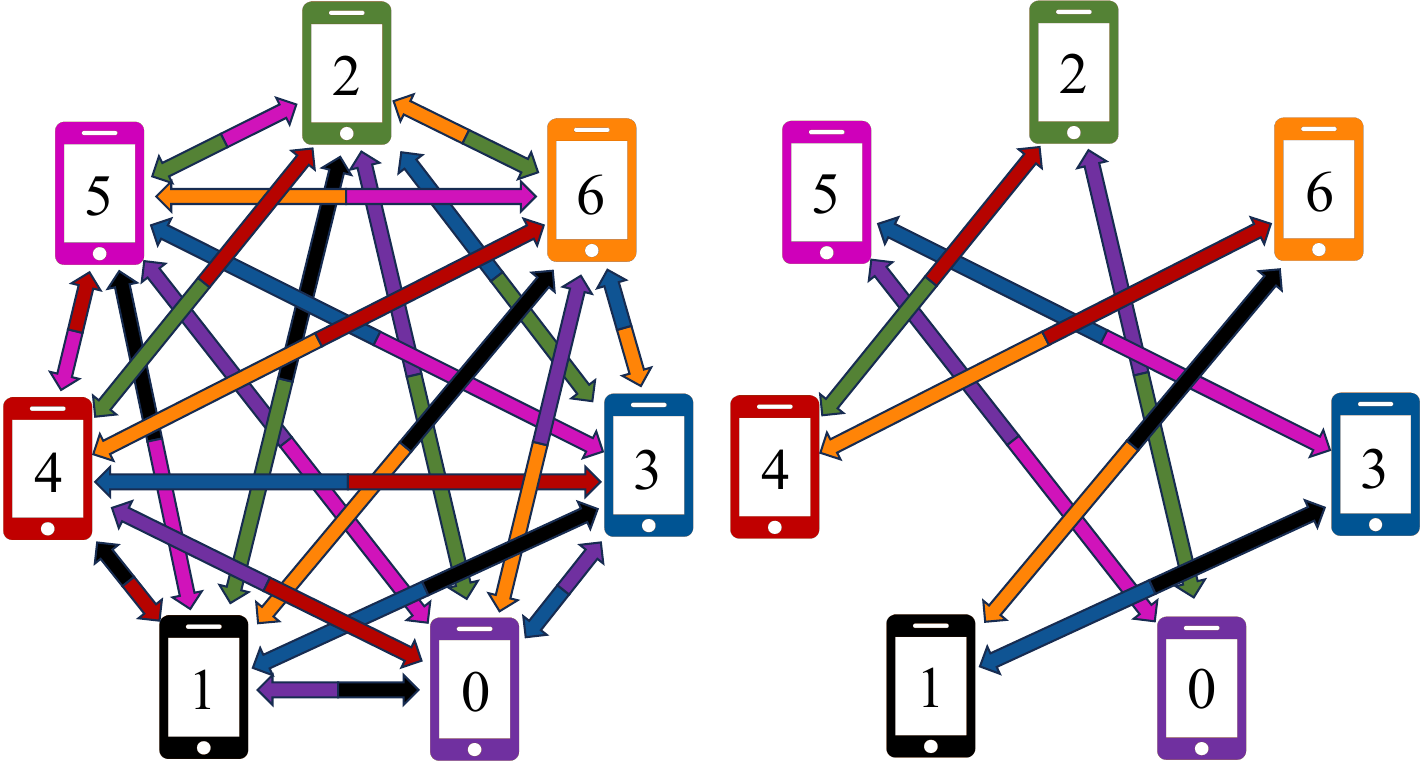}   
  \caption{SecAgg (left) vs. CESA (right): pair selection and shared secret creation.}
  \label{fig:SecAgg_vs_CESA} \vspace{-6mm}
\end{figure}

In recent years, Federated Learning (FL) \cite{mcmahan2017communication} has emerged as a decentralized and privacy-preserving training method, particularly designed for sensitive client data. In FL, models are trained on local devices, and a central server aggregates the gradients to build a new global model. Hence, the server does not have direct access to raw data. However, FL is vulnerable to security risks such as model inversion attacks where an \textbf{honest-but-curious curious server follows the aggregation rules but tries to infer sensitive client data from trained models}. Google's Secure Aggregation protocol (SecAgg) \cite{bonawitz2017practical} enhances data privacy by enabling clients to mask their trained models by using a double masking technique that combines the client's trained model with masks generated from individual random element and shared secrets for every pair of clients (Fig. \ref{fig:comparison_fl_secagg}). While SecAgg mitigates the privacy issue, it considerably increases communication and computation costs that grow with the number of clients. Building upon SecAgg, this poster proposes an efficient secure aggregation protocol called CESA with communication complexity independent of the network size (Fig. \ref{fig:SecAgg_vs_CESA}). Compared to SecAgg, CESA 1) eliminates the need for using a mask generated from a random element, 2) does not apply any encryption, and 3) creates only \textbf{two shared masks per client} by using public keys of two pairs. This approach reduces communication and computation costs while avoiding model inversion attacks in honest-but-curious scenarios. CESA is designed for networks with low delay variations and limited client dropouts.

\section{Key Agreement Algorithm}
\vspace{-0.15mm}

This section explains the key agreement algorithm that is used in secure aggregation. The process begins with two publicly known parameters: a large prime $p$ and a primitive root $g$ modulo $p$. Each client $i$ generates a private key $a_i$ randomly chosen from $[1, p-1]$ and computes a corresponding public key $A_i \equiv g^{a_i} \pmod{p}$. Similarly, client $j$ generates a private key $a_j$ and the public key $A_j \equiv g^{a_j} \pmod{p}$. Then, clients $i$ and $j$ exchange their public keys and compute shared secrets. Client $i$ computes the shared secret $s_{i,j} \equiv A_j^{a_i} \equiv (g^{a_j})^{a_i} \equiv g^{a_i a_j} \pmod{p}$, and client $j$ computes the shared secret $s_{j,i} \equiv A_i^{a_j} \equiv (g^{a_i})^{a_j} \equiv g^{a_i a_j} \pmod{p}$. Hence, $s_{i,j} = s_{j,i}$ and these two clients create the same shared secret.

\section{Background}
\vspace{-0.15mm}

 The section explains the SecAgg protocol. We denote the number of training rounds, the client index in the sorted participating list, and the number of participating clients as $n$, $i$, and $|C|$, respectively. We assume there is no client dropout in the following calculation of the number of exchanged messages. (1) \textbf{Broadcasting the global model:} The server broadcasts the initial global model to clients. Each client ${i}$ generates \textbf{two} private-public key pairs as $(SK_{i}^1, PK_{i}^1)$ and $(SK_{i}^2, PK_{i}^2)$. Then, it sends its public keys to the server. (3) \textbf{Broadcasting public keys:} The server broadcasts public keys to all clients. (4) \textbf{Client-side preparation:} Each client $i \in C$ generates a random element $b_{i}$, then divides $b_{i}$ and $SK_{i}^1$ into $|C|$ parts and assigns each part for every client $j \in C$ ($b_{i,j}$, $SK_{i,j}^1$). Then, Client $i$ encrypts a message $(i || j || b_{i,j} || SK_{i,j}^1)$ for each pair $j$ (by using a key generated from $SK_{i}^2$ and $PK_{j}^2$) to create a cipher text $e_{i,j}$. Finally, the client sends all generated $e_{i,j}$ to the server. (5) \textbf{Distribution of cipher texts:} The server collects these cipher texts and puts participated clients in $C_{1}$ set. Here, we assume that $|C_{1}| = |C|$, hence the server sends $(|C| - 1)$ encrypted values to every client. (6) \textbf{Masked model generation:} Each client $i$ creates $(|C| - 1)$ shared secrets with every other client $j$ by using $SK_{i}^1$ and $PK_{j}^1$. Then, client $i$ expands these created shared secrets and its random element $b_{i}$ by pseudo-random generator function $PRG$ to create an individual mask $M_{i}$ and a shared mask $M_{i,j} (\forall j \in C_{1})$. By using these masks, client $i$ computes the masked model $W_{i}^{mask}$ from its trained model $W_{i}$, which is sent to the server. (7) \textbf{Participants awareness:} The server creates a set $C_{2}$ from clients that sent their masked models. Then, the server sends the set $C_{2}$ to all the clients $j \in C_{2}$. Then, each client $i$ understands about participants and decrypts the received encrypted values $e_{j,i}$ by using a key generated from $SK_{i}^2$, and $PK_{j}^2$. Thus, the client achieves $b_{j,i} (\forall j \in C_{2}$) and sends them to the server. (8) \textbf{Global model aggregation:} The server gathers $(|C| - 1)$ portions of random elements of participants in $C_{2}$ and reconstructs $b_{j} (\forall j \in C_{2})$, then expands it by $PRG$ to generate individual mask $M_{j} (\forall j \in C_{2})$. Finally, it aggregates the global model by $\sum_{i \in \{C_{2}\}} W_{i}^{mask} - \sum_{i \in \{C_{2}\}} M_{i}$. Based on \cite{bonawitz2017practical}, the communication cost of each client is $O(|C|)$, and for the server is $O(|C|^2)$.

\section{The Proposed Method}
\vspace{-0.25mm}

This section introduces the overall FL embedded in CESA and theoretically analyzes the communication cost of each method.
The process of messages passing in CESA can be categorized into three main phases. \textbf{Phase $I$} (Initialization): The server \textbf{broadcasts} the initial model to all clients. Simultaneously, all clients receive common public parameters from a trusted third party. Then, each client generates a unique public-private key pair from the public parameters and sends the public key to the server. Upon collecting all public keys, the server broadcasts a set of all public keys. \textbf{Phase $II$} (Shared mask generation at training round 1): Upon receiving the public keys, each client calculates the index of two other clients in the participating list to create shared secrets using their public keys. Client $i$ calculates the index of its pairs as $FP_{i} = [(i+offset) \mod |C|]$ and $SP_{i} = [(i-offset+|C|) \mod |C|]$. Here, $offset$ is a random integer within the range of $[2, \left\lfloor \frac{|C|-1}{2} \right\rfloor]$ (where $|C| \geq 7$). After finding the pairs, the client generates shared secrets, which are used in a pseudo-random number generator that creates two shared masks (denoted as $M_{i, FP_{i}}$ and $M_{i, SP_{i}}$). \textbf{Phase $III$} (from training round 2): Each client $i$ performs model training and then computes the masked model as $W_{i}^{mask} = W_{i} + M_{i, FP_{i}} + M_{i, SP_{i}}$, where $W$ is the trained model. The computed $W_{i}^{mask}$ is sent to the server. Then, the server generates the new global model by aggregating all masked models. The masks cancel out each other due to the pairwise generation of shared secrets, and since $M_{i, FP_{i}} = - M_{FP_{i}, i}$ the sum of masked models equals the sum of unmasked trained models. At the end of the aggregation process, the server broadcasts the new global model to all clients. Considering all communications after $n$ FL rounds, the total number of messages sent from all clients are $(n + 1) \times |C|$, and from the server are $n + 1$ messages.

\section{Evaluation}
\vspace{-0.15mm}

\begin{figure}[t]
    \centering
    \begin{subfigure}{0.22\textwidth}
       \centering
        \includegraphics[width=30 mm, height=30 mm]{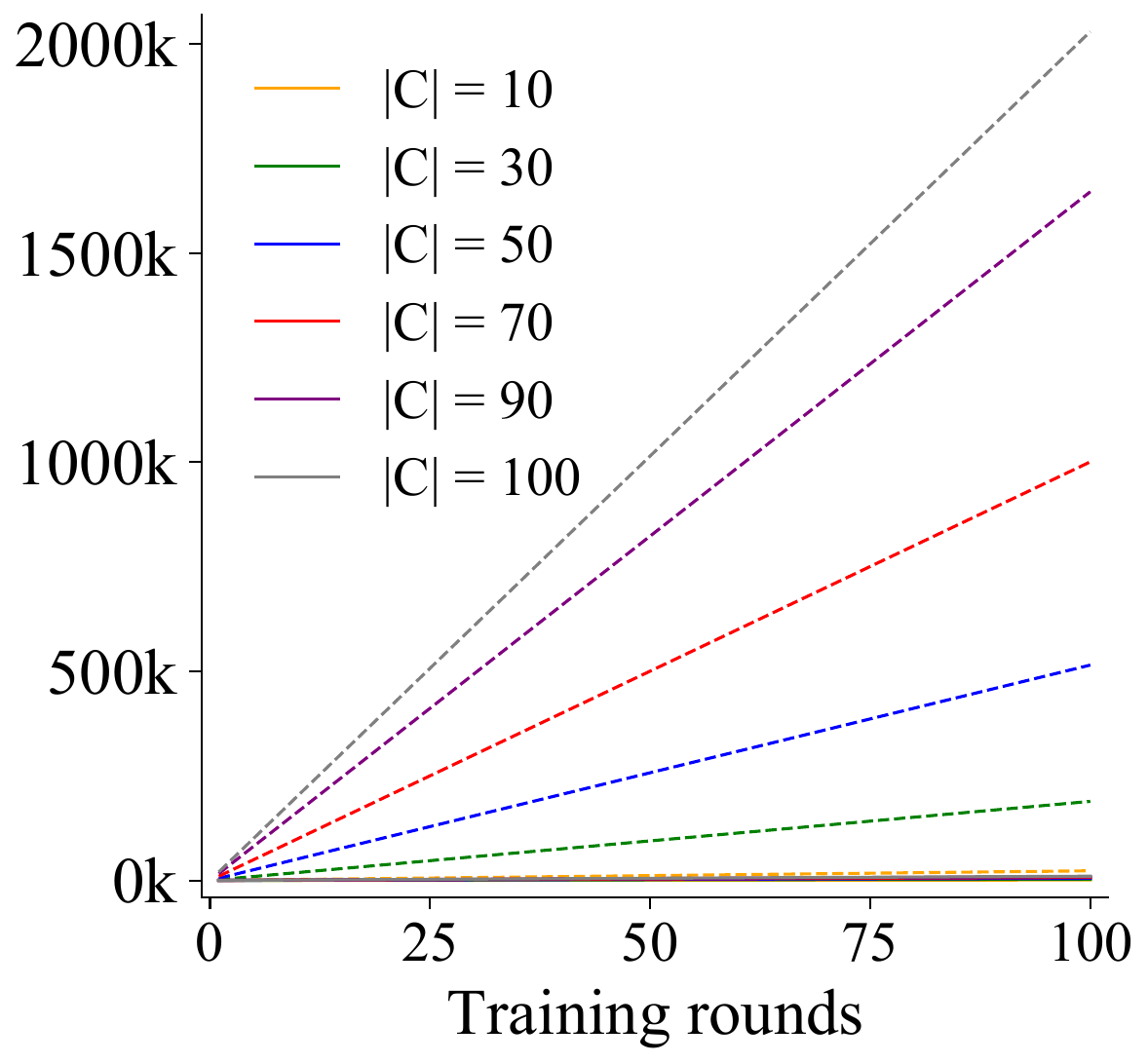}
        \caption{From clients to the server.}
        \label{fig:client_data} \vspace{-2mm}
    \end{subfigure}
    \hspace{1mm}
    \begin{subfigure}{0.22\textwidth}
    	\centering
        \includegraphics[width=30 mm, height=30 mm]{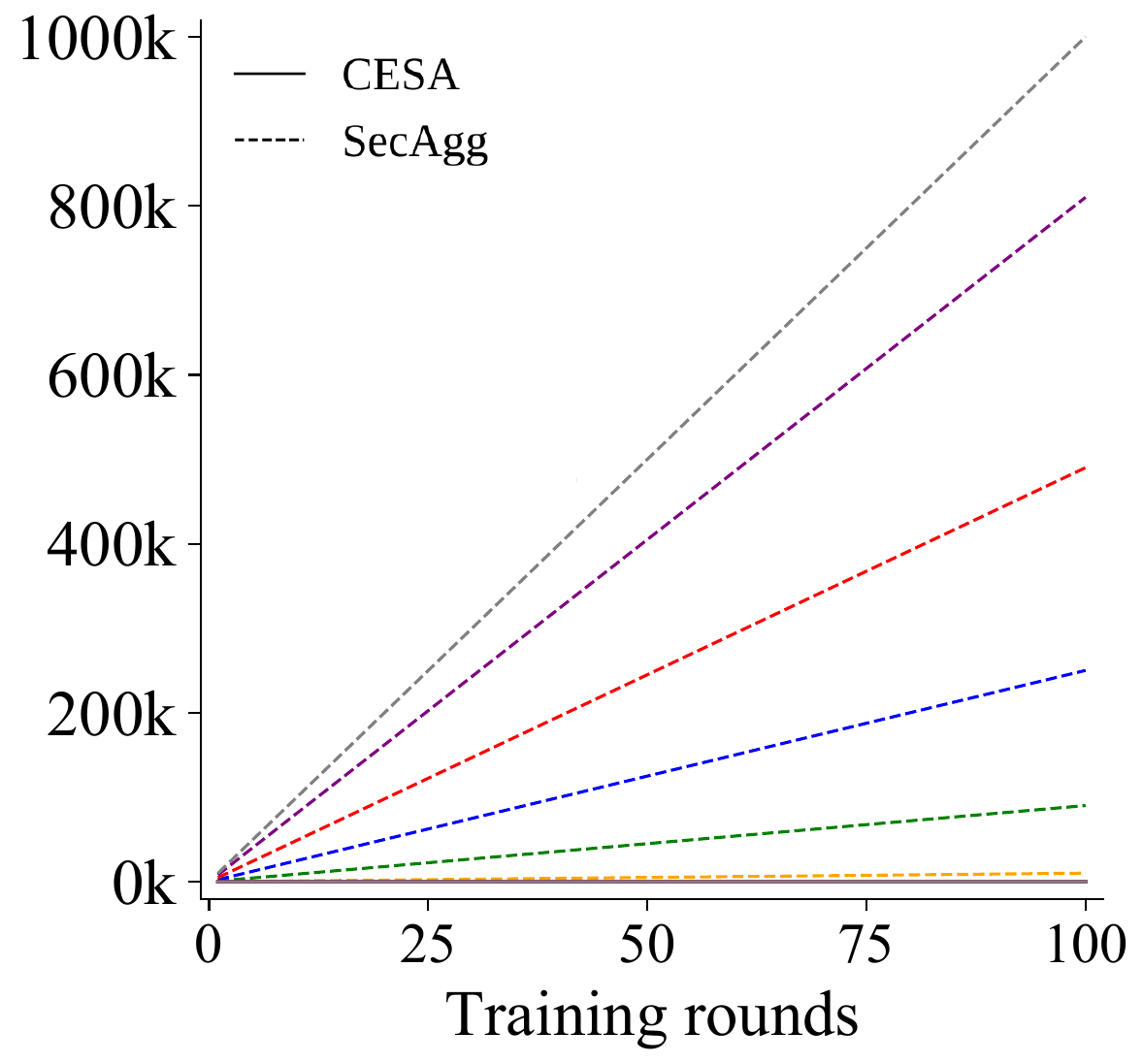} 
        \caption{From the server to clients.}
        \label{fig:server_data} \vspace{-2mm}
    \end{subfigure}
    \caption{Transmitted messages between clients and the server.}
    \label{fig:mainmetrics}  \vspace{-4mm}
\end{figure}

This section evaluates the communication efficiency of CESA compared to SecAgg by calculating the number of messages transmitted between clients and the server over $100$ rounds for varying numbers of clients (Fig. \ref{fig:mainmetrics}). In this figure, except for broadcast messages, each message carries only 1 value. As shown in Fig. \ref{fig:mainmetrics}, CESA maintains a constant number of messages sent from the server to clients, independent of client count. In contrast, SecAgg's server-side message count increases in more extensive networks, leading to higher communication demands.

\section{Conclusion}
\vspace{-0.15mm}

In this poster, we presented an FL's communication-efficient secure aggregation protocol in honest-but-curious scenarios for stable networks with limited client dropouts. Unlike SecAgg, which requires dividing and encrypting keys, generating shared masks for every other client, and computing individual masks, CESA computes shared masks only two pairs per client without performing any encryption. Our initial evaluation revealed that CESA significantly reduces the communication cost of SecAgg and provides a defense mechanism against the model inversion attack.

\bibliography{citation.bib}
\bibliographystyle{ieeetr}


\end{document}